\documentclass[twocolumn]{aastex6}

\shorttitle{Worlds Without Moons}
\shortauthors{Stephen R. Kane}

\begin{document}

\title{Worlds Without Moons: Exomoon Constraints for Compact Planetary
  Systems}

\author{Stephen R. Kane}
\affil{Department of Physics \& Astronomy, San Francisco State
  University, 1600 Holloway Avenue, San Francisco, CA 94132, USA}
\email{skane@sfsu.edu}

%%%%%%%%%%%%%%%%%%%%%%%%%%%%%%%%%%%%%%%%%%%%%%%%%%%%%%%%%%%%%%%%%%%%

\begin{abstract}

One of the primary surprises of exoplanet detections has been the
discovery of compact planetary systems, whereby numerous planets
reside within $\sim$0.5~AU of the host star. Many of these kinds of
systems have been discovered in recent years, indicating that they are
a fairly common orbital architecture. Of particular interest are those
systems for which the host star is low-mass, thus potentially enabling
one or more of the planets to lie within the Habitable Zone of the
host star. One of the contributors to the habitability of the Earth is
the presence of a substantial moon whose tidal effects can stabilize
axial tilt variations and increase the rate of tidal pool
formation. Here we explore the constraints on the presence of moons
for planets in compact systems based on Hill radii and Roche limit
considerations. We apply these constraints to the TRAPPIST-1 system
and demonstrate that most of the planets are very likely to be worlds
without moons.

\end{abstract}


%%%%%%%%%%%%%%%%%%%%%%%%%%%%%%%%%%%%%%%%%%%%%%%%%%%%%%%%%%%%%%%%%%%%

\section{Introduction}
\label{intro}

Over the past few decades, the field of exoplanets has provided no
shortage of orbital architectures for which we have no analog in our
own planetary system. These include hot Jupiters, massive planets in
eccentric orbits, circumbinary planets, and compact planetary
systems. The initial windfall of compact planetary systems came from
the early results of the {\it Kepler} mission. The bias of the transit
method towards short orbital periods \citep{kan08} meant that these
compact systems exist in the region of parameter-space where {\it
  Kepler} had the strongest sensitivity. However, the frequently faint
host stars also meant that masses could not easily be measured from
precision radial velocities. Fortunately the short period of the
compact system planets and the precision of the {\it Kepler}
photometry allowed the measurement of the masses through Transit
Timing Variations (TTVs) \citep{mir02,ago05,hol05}. The measurements
of the planetary masses allow, amongst other things, exploration of
the orbital stability of these systems
\citep{ray09,fun10,zha13,han14,bec17}.

Notable examples of compact planetary systems with measured masses are
Kepler-11 \citep{lis11,lis13} and Kepler-80 \citep{xie13,mac16}. The
recently discovered TRAPPIST-1 system provides a compact orbital
architecture where some of the planets appear to occupy the Habitable
Zone (HZ) of the host star. Initial observations uncovered three
terrestrial planets transiting the faint ultracool dwarf star
\citep{gil16}. Continued observations using Spitzer revealed an
additional four terrestrial transiting planets, including TTVs that
allowed the masses of the planets to be determined \citep{gil17}. A
longer time baseline of photometry provided by the {\it K2} extension
of the {\it Kepler} mission confirmed the orbital period of the outer
planet that placed the orbit near the TRAPPIST-1 snow line
\citep{lug17}.

Much has been written about the detection of exomoons and their
potential habitability
\citep{rey87,wil97,wei10,for13,hin13,hel13,hel14b,kip15}. It is also
commonly held that the presence of a moon with substantial mass played
a key role in Earth's habitability through obliquity
stabilization. Early work by \citet{las93} indicated that a moonless
Earth would have extreme variations in obliquity resulting in dramatic
climate changes. A study by \citet{tom96} expanded the chaotic
obliquity calculations by included the effects of the tidal expansion
of the moon. Further simulations by \citet{lis12} and \citet{li14}
demonstrated that the Moon does indeed stabilize the Earth's
obliquity, though not at the previously determined amplitude and thus
a moonless Earth does not necessarily preclude habitability.

Here we explore the role of orbital architecture and planetary masses
and radii on the ability of the planets to harbor exomoons. We show
that gravitational influence of planets in compact planetary systems
places severe constraints on the presence of exomoons and we use the
TRAPPIST-1 system as an example. In Section~\ref{radii}, we quantify
the limits imposed on exomoon orbits by the Hill radius and Roche
limits of a planet. In Section~\ref{con}, we construct analytical
limits for exomoon stability regions in the context of compact
planetary system architectures. Section~\ref{t1} applies these
constraints to the TRAPPIST-1 system and demonstrates why many of the
TRAPPIST-1 planets cannot harbor moons. We provide discussion and
concluding remarks in Section~\ref{conclusions}.

%%%%%%%%%%%%%%%%%%%%%%%%%%%%%%%%%%%%%%%%%%%%%%%%%%%%%%%%%%%%%%%%%%%%

\section{Hill Radius and Roche Limit}
\label{radii}

The two main considerations for the region in which a moon can be
maintained in orbit around a planet are the Hill radius, which defines
the outer limit of the region, and the Roche limit, which defines the
inner limit \citep{kip09,hel12}. The Hill radius, $R_H$, is given by
\begin{equation}
  R_H = a_p \left( \frac{M_p}{3 M_\star} \right)^{\frac{1}{3}}
  \label{rh}
\end{equation}
where $a_p$ is the planetary semi-major axis, $M_p$ is the planetary
mass, and $M_\star$ is the stellar mass. The Roche limit, $R_R$, is
given by
\begin{equation}
  R_R \simeq 2.44 R_p \left( \frac{\rho_p}{\rho_m}
  \right)^{\frac{1}{3}}
  \label{rr}
\end{equation}
where $R_p$ is the planetary radius, $\rho_p$ is the planetary
density, and $\rho_m$ is the moon density. For example, the Hill
radius of the Earth is 0.01~AU and the Roche limit is $1.26 \times
10^{-4}$~AU, a factor of $\sim$78.5 difference. For an Earth-analog in
a compact system with a semi-major axis of 0.05~AU, the Hill radius
shrinks to $6 \times 10^{-4}$~AU which results in a factor of
$\sim$4.5 difference with the Roche limit.

Given that the radius of the Hill sphere is a theoretical
approximation, the practical outer limit of a stable orbit for a moon
can be significantly smaller due other perturbation effects, such as
those originating from the host star. As such, the outer limit may be
described as $\chi R_H$ where $\chi$ is a reduction factor due to the
above described effects \citep{hol99}. Further perturbation sources
are other bodies in the system, particularly planetary bodies
\citep{gon13,pay13}. Furthermore, formation processes play a role in
the destabilization of moons, such as their potential for resonant
removal during planetary migration \citep{spa16}. For example,
\citet{bar02} and \citet{kip09} estimate $\chi \approx 1/3$,
effectively reducing the outer limit of a stable moon orbit to $1/3
\times R_H$. An additional process that will reduce the size of the
allowed region for long-term moon orbital stability is the evolution
of the orbit through tidal interactions with the planet \citep{sas14},
effectively resulting in an exchange of angular momentum. The inward
or outward direction of the moon migration depends upon the ratio of
the orbital period to the rotation period of the planet \citep{bar02}
and also if the orbit is prograde or retrograde \citep{dom06}.  Note
that the semi-major axis of the Moon's orbit is $\sim$26\% of the
Earth's Hill radius, which is close to the $\chi \approx 1/3$ criteria
mentioned above. For comparison, the outermost of the Galilean moons,
Callisto, has a semi-major axis that is $\sim$3\% of Jupiter's Hill
radius. Clearly the effective size of the Hill sphere for planets in
compact systems can be greatly reduced by the properties of the local
environment.

%%%%%%%%%%%%%%%%%%%%%%%%%%%%%%%%%%%%%%%%%%%%%%%%%%%%%%%%%%%%%%%%%%%%

\section{Exomoon Constraints for Compact Planetary Systems}
\label{con}

\begin{figure}
  \includegraphics[angle=270,width=8.5cm]{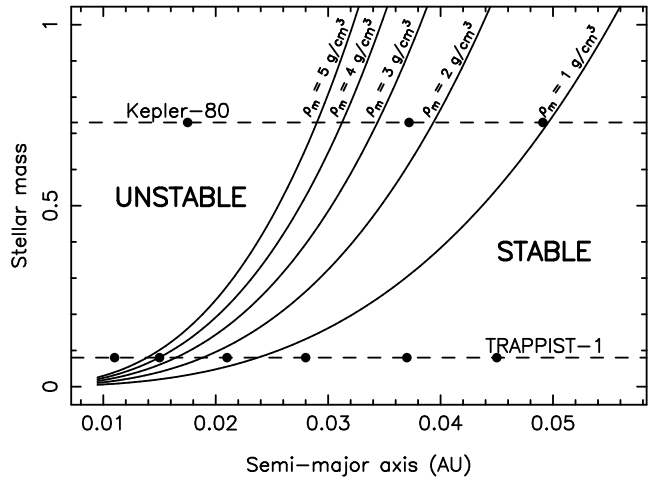}
  \caption{The locations of equality between the Hill radius and Roche
    limit (see Equation~\ref{limit}) for five different mean densities
    of a potential moon. Stable moon orbits cannot exist above the
    curve for a given density. For the purposes of these calculations,
    we adopt a reduction factor of $\chi = 1/3$. The horizontal dashed
    lines indicate the stellar masses for two example compact
    planetary systems, Kepler-80 and TRAPPIST-1, with the locations of
    the planets shown on each line.}
  \label{equality}
\end{figure}

Here we consider the Hill and Roche limit radii described in
Section~\ref{radii} in the extreme environments of compact planetary
systems. Within the orbital regime of these systems, the gravitational
perturbations are dominated by the host star rather than the planets,
enabling stability through the shrinking of the planetary Hill spheres
\citep{nam10}. An upper limit of the presence of moons may be
determined by evaluated conditions where $\chi R_H = R_R$. Using
Equations \ref{rh} and \ref{rr}, we find the following relationship:
\begin{equation}
  \chi a_p \left( \frac{M_p}{3 M_\star} \right)^{\frac{1}{3}} = 2.44
  R_p \left( \frac{1}{\rho_m} \frac{M_p}{\frac{4}{3} \pi R_p^3}
  \right)^{\frac{1}{3}}
\end{equation}
which is simplified as
\begin{equation}
  \rho_m = 10.4 \frac{M_\star}{(\chi a_p)^3}
\end{equation}
A more convenient form of the relationship is as follows:
\begin{equation}
  \left( \frac{\rho_m}{\mathrm{g/cm}^3} \right) = 6.18 \times 10^{-6}
  \chi^{-3} \left( \frac{M_\star}{M_\odot} \right) \left(
  \frac{a_p}{\mathrm{1 \ AU}} \right)^{-3}
  \label{limit}
\end{equation}
Equation~\ref{limit} is plotted in Figure~\ref{equality} for densities
ranging from 1~g/cm$^3$ to 5~g/cm$^3$, where we have assumed a Hill
radius reduction factor of $\chi = 1/3$. Horizontal dashed lines
indicate the stellar masses for the Kepler-80 (0.73~$M_\odot$) and
TRAPPIST-1 (0.0802~$M_\odot$) host stars, the planets of which largely
fall within the range of semi-major axes shown
\citep{mac16,gil17}. Since the solid curves represent the boundary
where $\chi R_H = R_R$, moons of the given density cannot exist within
regions of parameter space that lie above the curve. For example, no
planet in the Kepler-80 system within 0.035~AU of the host star can
ever host a moon with a density $\leq 3$~g/cm$^3$. The cubic nature of
the relationship described in Equation~\ref{limit} does ensure that
the concise exclusion of moons predominantly affects those planets in
the very small semi-major axis regime. An interesting possibility is
that the moon constraints for compact systems may still allow the
presence of ring systems \citep{bar04,zul15}, particularly considering
the relatively long persistence of debris disks around low-mass stars
\citep{pla09}. Such rings around terrestrial planets are likely to
have low optical depth and not readily detectable from transit
photometry when they are coplanar with the inclination of the
planetary orbits.

%%%%%%%%%%%%%%%%%%%%%%%%%%%%%%%%%%%%%%%%%%%%%%%%%%%%%%%%%%%%%%%%%%%%

\begin{deluxetable*}{lcccccc}
  \tablecolumns{7}
  \tablewidth{0pc}
  \tablecaption{\label{t1tab} TRAPPIST-1 planetary Hill radii and
    Roche limits}
  \tablehead{
    \colhead{Planet} &
    \colhead{$M_p$ ($M_\oplus$)} &
    \colhead{$R_p$ ($R_\oplus$)} &
    \colhead{$a$ (AU)} &
    \colhead{$R_H$ ($10^{-3}$ AU)} &
    \colhead{$R_R$ ($10^{-3}$ AU)} &
    \colhead{$R_H/R_R$}
  }
  \startdata
  TRAPPIST-1 b & 0.85 & 1.086 & 0.011 & 0.244 & 0.120 &  2.04 \\
  TRAPPIST-1 c & 1.38 & 1.056 & 0.015 & 0.393 & 0.141 &  2.79 \\
  TRAPPIST-1 d & 0.41 & 0.772 & 0.021 & 0.370 & 0.094 &  3.94 \\
  TRAPPIST-1 e & 0.62 & 0.918 & 0.028 & 0.557 & 0.108 &  5.17 \\
  TRAPPIST-1 f & 0.68 & 1.045 & 0.037 & 0.756 & 0.111 &  6.80 \\
  TRAPPIST-1 g & 1.34 & 1.127 & 0.045 & 1.154 & 0.139 &  8.28 \\
  TRAPPIST-1 h & 0.31 & 0.715 & 0.060 & 0.936 & 0.086 & 10.86
  \enddata
  \tablecomments{Planetary masses, radii, and semi-major axes are from
    \citet{gil17}.}
\end{deluxetable*}

\section{No Moons in the TRAPPIST-1 System?}
\label{t1}

The TRAPPIST-1 system is an exceptional case of compact planetary
systems due to the extremely low mass of the host star. The
combination of transits and TTVs provide radii and masses for the
planets, allowing us to calculate the Hill radii and Roche limits for
each planet (see Section~\ref{radii}). Currently the outer planet
(planet h) does not have a reliable mass estimate, and so we used the
mean density of the other six planets of $\sim$0.84~$\rho_\oplus$ to
determine a mass estimate of 0.31~$M_\oplus$. The resulting
calculations for all seven planets are shown in
Table~\ref{t1tab}. Also shown are the ratios of the Hill radii and
Roche limits, where a Hill radius reduction factor of unity has been
used and a moon density of 3~g/cm$^3$ adopted for the Roche limit. For
comparison, the mean density of the Galilean moons is 2.6~g/cm$^3$ and
the mean density of the Earth's moon is 3.3~g/cm$^3$.  The ratios in
the final column of Table~\ref{t1tab} show that, even for the case of
$\chi = 1$, the difference between the Hill radii and Roche limits for
the inner planets is remarkably small, similar to the case for the
Earth-analog at 0.05~AU described in Section~\ref{radii}.

\begin{figure}
  \includegraphics[angle=270,width=8.5cm]{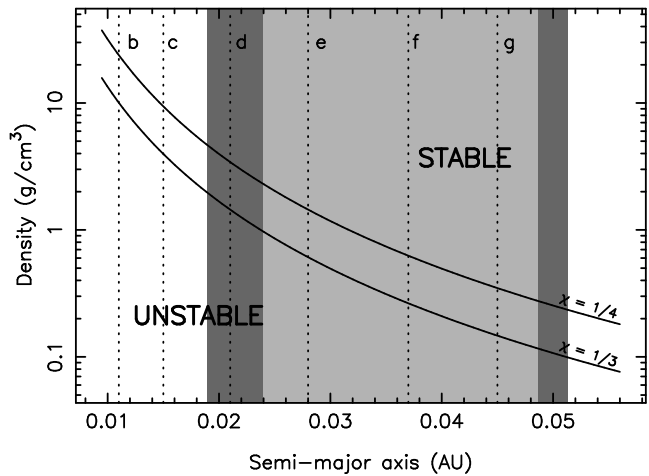}
  \caption{The allowed moon density as a function of semi-major axis
    for the TRAPPIST-1 system and for Hill radii reduction factors of
    $\chi = 1/3$ and $\chi = 1/4$. Stable moon orbits cannot exist
    below each of the curves. The vertical dotted lines indicate the
    location of the six innermost TRAPPIST-1 planets. The light-gray
    shaded region shows the extent of the conservative HZ, and the
    dark-gray shaded region represents the optimistic extension to the
    HZ.}
  \label{t1fig}
\end{figure}

Shown in Figure~\ref{t1fig} are moon exclusion boundaries (solid
lines) for the TRAPPIST-1 system as a function of mean density and
semi-major axis, with vertical dotted lines indicating the location of
the six innermost planets. The shaded regions show the extent of the
HZ for TRAPPIST-1, where we used the TRAPPIST-1 luminosity
(0.000524~$L_\odot$) and effective temperature (2559~K) provided by
\citet{gil17}, along with the HZ boundary equations of
\citet{kop13,kop14}. The shaded regions correspond to the
``conservative'' (light-gray, 0.024--0.049~AU) and ``optimistic''
(dark-gray, 0.019--0.051~AU) HZ boundaries, the definitions of which
depends upon assumptions regarding the longevity of liquid water on
the surfaces of Venus and Mars. Although the full extent of the
perturbations to potential moons is not quantified, the existence of
resonant chains of planetary orbits \citep{gil17} is expected to
increase the amplitude of such perturbations. Therefore, we have used
Hill radii reduction factors of $\chi = 1/3$ and $\chi = 1/4$ in the
figure. In this case, the exclusion boundaries imply that the
parameter space below the curve show where stable orbits of moons are
not possible. For example, the $\chi = 1/4$ curve indicates that
planets b--e cannot host moons of reasonable ($\lesssim 3$~g/cm$^3$)
mean densities. As noted at the end of Section~\ref{radii}, the Hill
radii reduction factors do not include the effects of moon orbital
migration due to a tidal exchange of angular momentum
\citep{sas12}. Thus the region for which each planet can sustain a
moon in a stable orbit over long time periods is further reduced such
that it is highly unlikely that any of the planets represented in
Figure~\ref{t1fig} host moons. Properties of the host star, such as
activity and rotation rate, suggest that the planetary system may be
relatively young \citep{gil16}. The outer planets may therefore
currently hold moons in temporarily stable locations although, as
noted earlier, the process of migration into regions of resonance
trapping may have stripped them already \citep{spa16}.

%%%%%%%%%%%%%%%%%%%%%%%%%%%%%%%%%%%%%%%%%%%%%%%%%%%%%%%%%%%%%%%%%%%%

\section{Conclusions}
\label{conclusions}

Although the discovery of compact planetary systems was initially a
surprising outcome of exoplanet observations, it now seems as though
they may be a quite common occurrence. In many cases, their
architectures have been compared to that of the Jovian system leading
to speculation of similar formation mechanisms \citep{kan13}. An
aspect of particular interest is the presence of compact systems
around low-mass stars where the chances of temperate terrestrial
planets are greatly enhanced. The TRAPPIST-1 system is an ideal
example of such a system, with three planets within the conservative
HZ and four within the optimistic HZ. This has resulted in the
postulated habitability of the TRAPPIST-1 planets \citep{gil17},
including detailed models of the planetary atmospheres
\citep{wol17}. The presence of moons can have a positive impact on
habitability, such as promoting tidal pools and stabilizing rotational
obliquity. Insofar as habitability is reduced by a lack of moons,
compact planetary systems such as TRAPPIST-1 may suffer due to the
constraints on the presence of moons outlined in this work. As shown,
a dominant deciding factor is the mean density of the moon which
depends on whether they formed in situ around the planet and the
location of planet formation. Planetary migration of planets near or
beyond the snow line may result in a lower than expected moon density
which will in turn ensure their removal according to the calculations
described herein.

The presence of moons, or lack thereof, in the system is a testable
hypothesis given the techniques available from transit variations. A
moon will produce TTV and/or Transit Duration Variation (TDV) effects
in the timing and shape of the transit photometry, depending on the
mass and separation of the moon \citep{sar99,kip09}. Additionally, the
size of the moon will play a major role in its detectability in the
photometric data. For example, the radius ratio of the Earth's moon
and the Earth is 0.272; thus the Earth's moon has a transit depth that
is 7.4\% as large as the Earth's. Such a deviation would have been
detected in the precision {\it Spitzer} photometry presented by
\citet{gil17}, and so moons of comparable size ratios are likely not
present, consistent with the findings of this work. Moons as small as
the minimum size to be round (radius of 200--300~kms) are unlikely to
be detected without exceptional circunstances, such as phase folding
planetary transits for a low-mass star with relatively low stellar
activity \citep{hel14a}. Long-term monitoring with current and future
facilities, such as thirty-meter class ground-based telescopes and the
James Webb Space Telescope, will place further constraints on the
presence of moons in these system.

%%%%%%%%%%%%%%%%%%%%%%%%%%%%%%%%%%%%%%%%%%%%%%%%%%%%%%%%%%%%%%%%%%%%

\section*{Acknowledgements}

The author would like to thank Duncan Forgan, Ren\'e Heller, and
Natalie Hinkel for their insightful feedback on this work. This
research has made use of the NASA Exoplanet Archive, which is operated
by the California Institute of Technology, under contract with the
National Aeronautics and Space Administration under the Exoplanet
Exploration Program. The results reported herein benefited from
collaborations and/or information exchange within NASA's Nexus for
Exoplanet System Science (NExSS) research coordination network
sponsored by NASA's Science Mission Directorate.

%%%%%%%%%%%%%%%%%%%%%%%%%%%%%%%%%%%%%%%%%%%%%%%%%%%%%%%%%%%%%%%%%%%%

\end{document}